# Extension of Spatial *k*-Anonymity:
New Metrics for Assessing the Anonymity of Geomasked Data Considering Realistic Attack Scenarios

Simon Cremer[1], Lydia Jehmlich[1], and Rainer Lenz[1,2]

[1]Institute for Production, Cologne University of technology, arts and sciences, 50679 Cologne, Germany

[2]Department of Statistics, Technical University of Dortmund, 44221 Dortmund, Germany

simon.cremer@th-koeln.de, lydia.jehmlich@th-koeln.de,
rainer.lenz@th-koeln.de

**Abstract.** Spatial data are gaining increasing importance in many areas of research. Particularly spatial health data are becoming increasingly important for medical research, for example, to better understand relationships between environmental factors and disease patterns. However, their use is often restricted by legal data protection regulations, since georeferenced personal information carries a high risk of re-identification of individuals. To address this issue, what are called geomasking methods are applied to guarantee data protection through targeted displacement of individual data points, while simultaneously maintaining analytical validity within a tolerable range.

In the current literature the degree of anonymity of such anonymized georeferenced datasets is often measured by the so-called metric of spatial *k*-anonymity. However, this metric has considerable shortcomings, particularly regarding its resilience against realistic data attack scenarios. This article classifies the potential data attack scenarios in the context of anonymized georeferenced microdata and introduces appropriate metrics that enable a comprehensive assessment of anonymity adapted to potential data attack scenarios.

# 1    Introduction

The purposeful use of health data is fundamental for medical progress, prevention, and improved patient care (Federal Ministry of Health, 2025). Moreover, public health data with spatial reference can be of immense value for research. For example, enabling the analysis of local disease phenomena, inequalities in access to health-related infrastructure, or regional differences in healthcare provision (Kistemann et al., 2011). According to the GDPR (Art. 9), such health data are considered particularly sensitive and hence there is need for data protection. The necessary level of data protection, however, may hinder the value-generating use of health data. In order to make health datasets usable, they must be anonymized – conventionally by aggregation. While data aggregation in most cases ensures a high degree of privacy, it comes at the expense of data utility: the more strongly the data are coarsened, the more their informational content and thus their analytical suitability decline. Balancing this trade-off is the aim of applying so-called geomasking methods (Armstrong, 1999), which are perturbing spatial data while simultaneously preserving their analytical validity to a required degree. In the context of health data, this specifically means publishing sensitive attributes – such as patients' health status – with as precise local occurrence as possible (geocoordinates) and without allowing re-identification of the individuals concerned.

But when is a dataset sufficiently anonymous? To describe the degree of anonymity of a geomasked dataset, the criterion of spatial $k$-anonymity is primarily used. A closer examination of this measure, however, reveals that it is not well-suited to withstand various conceivable data attack scenarios. In the current literature, such data attack scenarios are generally not sufficiently considered. This may be one reason why geomasking methods have so far found little application in the day-to-day practices of public and private data holders. To ensure the anonymity of some geomasked dataset, appropriate metrics are needed enabling the decision maker to reliably evaluate the associated anonymity. For this purpose, realistic attack scenarios are systematized. That is, scenarios that could be undertaken by a potential data intruder targeting to reveal confidential information of previously geomasked data. Building on this, we propose metrics that are capable of adequately measuring the level of privacy protection, even if such scenarios occur.

In the following, we assume a health dataset that contains at least the geocoordinates of individuals' residential addresses as well as at least one sensitive analytical attribute (e.g., socio-demographic information or ICD code).

# 2    Framework of a data attack

A disclosure of confidential information occurs when a person or organization gains knowledge about another person or organization that was previously unknown to them by using released (anonymized) data. Two types of disclosure risk are distinguished: identity disclosure and attribute disclosure. Identity disclosure occurs when a respondent's identity can be associated with a disseminated data record containing confidential information (e.g. see Duncan et al., 2001). Attribute disclosure occurs when either an attribute value contained in the disseminated data or an estimated attribute value derived from it can be associated with the respondent (ibid.).

Within a data attack, a potential data intruder attempts to match external information with the confidential data, which we refer to as external data and target data. For this purpose, an intruder relies on variables that the external data and the target data share, the so-called key variables, or more generally, the quasi-identifiers. In addition, it makes sense to highlight two particular main strategies of data attack. The first is the so-called match for a single individual, where a potential data intruder aims to uncover a single person or information about that person within an anonymized dataset. The second scenario is the so-called database cross match, where the objective is to uncover as many individuals as possible within an anonymized dataset in order to enrich an existing external dataset (Lenz et al., 2006; Lenz, 2006).

The following section addresses the additional knowledge potentially available to a data intruder and the types of variables that are relevant in the context of a data attack.

## 2.1 Additional knowledge and types of variables

The additional knowledge of a data intruder broadly consists of an external database, combined with additional meta-information that allows the external data to be successfully linked with the confidential target data.
To later formulate and analyze specific attack scenarios, it is first necessary to examine in more detail the variables contained in both external and target data.

**Direct identifiers**. A direct identifier is a variable that appears in both the external and the target data and that can uniquely link a record from the external data to the corresponding individual in the target data. Examples include name and date of birth, a personal email address, or a registration number from national official statistics. Likewise, the residential address or geocoordinates of a person's residence may in the worst-case lead directly to identification, since there are addresses at which only one single individual is registered.

**Quasi-identifiers**. These variables can, often in combination, enable the identification of an individual (identity disclosure). For example, if a data intruder knows the age and gender of an individual who participated in a survey, and these variables were also included in the anonymized dataset, they may be used either to uniquely identify the individual or at least to narrow down the set of potential candidates in the target dataset. Consequently, all such variables in the target dataset that could be available to a potential data intruder are considered as quasi-identifiers. This knowledge may stem from publicly available sources such as population statistics, social networks, geographic maps, or other databases. It may also include personal information obtained through social contacts, such as from a neighborhood or workplace. Typically, quasi-identifiers are modified during the anonymization process of the target data in order to prevent or at least complicate accurate linkage.

**Sensitive attributes**. These variables refer to sensitive information associated with an individual in the target data that must be protected from disclosure by potential data intruders, for instance such as a rare disease. Additional variables like age, gender, income, occupation, length of hospital stay, or comorbidities may also qualify as sensitive attributes. In general, data holders should classify all non-public variables in a dataset as confidential and make every effort to prevent disclosure of sensitive information about the individuals contained. In the context of specific data attack scenarios, it may even be reasonable to modify sensitive attributes during the process of anonymization. The goal here is that even if an individual were re-identified, just minimal information could be inferred from these attributes.

In this article, particular focus is placed on address data. Interestingly, geocoordinates may – depending on the perspective of a data attack (see section 2.2 below) – be used both as (quasi-)identifiers and sensitive attributes.

So far, we have already described the most important component of a potential data intruder's additional knowledge: an external dataset containing the objects of the intruder's interest along with a set of quasi-identifiers. Moreover, the presence or absence of the following auxiliary information is characteristic for an attack scenario and decisive for a successful disclosure of sensitive information:

- **Knowledge on the geomasking method(s) applied:** A data intruder may be informed about the functioning of the applied methods and/or the associated parameters used. Quite often, the methodology is described in the metadata of data products being published by data holders – among other reasons, to allow scientific data users to better interpret the analyses conducted.

- **Knowledge on the participation of individuals:** A data intruder possesses what is called participation knowledge if he or she has information on whether or not individuals or households of interest took part in the survey. Such knowledge may range from a single individual to all individuals or households included in the target dataset. In the case of a complete survey (according to a population), knowledge on participation is trivially assumed.

Specifically for the attack scenarios on geodata presented below, participation knowledge implicitly includes that the external dataset contains the original residential addresses of one or more individuals within the target dataset.

## 2.2 Perspectives of data attack

In the following, we define data attack perspectives that differentiate according to the 'direction' of the attack and the availability of auxiliary knowledge on participation (yes/no) or methods applied (yes/no). As mentioned previously, in the current paper, knowledge on participation includes that a potential data intruder not only knows whether the individuals of interest participated in the dataset but also has access to their actual residential addresses.

The term direction is understood to describe whether an intruder, on the one hand, attempts to assign an anonymized data point to its corresponding original data point (perspective 1: original address as target attribute), which means a sort of identity disclosure. Or, on the other hand, whether the intruder already possesses the original addresses of the individuals of interest and seeks to link them with the corresponding anonymized ones (perspective 2: original address as quasi-identifier), which means a sort of attribute disclosure.

To describe the different data attack perspectives and the derived anonymity metrics, we require some basic notations, which are listed in bullet-point form below.

- **Study area**
  - $U \subset \mathbb{R}^2$
    Represents the two-dimensional area of the region which is studied (e.g., the city area of Cologne)

- **Real addresses**
  - $B = \{b_1, b_2, \ldots, b_m\} \subset U$
  - Discrete set of all real existing residential addresses in the study area

- **Target data**
  - $P = \{p_1, p_2, \ldots, p_n\}$
  - Describes the non-anonymized confidential target data (persons)

- **Anonymized target data**
  - $P' = \{p_1', p_2', \ldots, p_n'\}$
  - Defines an anonymized dataset derived from $P$

- **Addresses associated with $P$ und $P'$**
  - Set of (original) addresses in the target data $P$:
    $A = \{a_1, a_2, \ldots, a_n\} \subseteq B$
    Each record $p_i \in P$ is associated with an address $a_i \in A$
  - Set of anonymized addresses in $P'$:
    $A' = \{a_1', a_2', \ldots, a_n'\} \subseteq B$
    Each record $p_i' \in P'$ is associated with an address $a_i' \in A'$

Note: In dependence of the anonymization method, it may hold $A' \subseteq B$. However, if addresses are perturbed randomly within the study area $U$, this inclusion does not necessarily hold

- **External data**
  - $Q = \{q_1, q_2, \dots, q_s\}$
  - This dataset contains at least, the residential address coordinates of the individuals of interest
  - Set of addresses in the external dataset:
    $A^q = \{a_1^q, a_2^q, \dots, a_s^q\}$
    Without knowledge of participation: $A^q \subseteq B$
    With knowledge of participation: $A^q \subseteq A$

- **Distance function**
  - The Euclidean norm is chosen as the distance function:
    $d: U \times U \longrightarrow \mathbb{R}_{\geq 0}, \quad d(x, y) = \| x - y \|_2$

- **Circle neighborhood of some point x**
  - $\mathrm{C}(x, r) := \{y \in U \mid d(x, y) \leq r\}$
  - Circle neighborhood of $x$ restricted to $D$:
    $\mathrm{C}_D(x, r) := \{y \in D \mid d(x, y) \leq r\}$

- **Method-specific displacement area**
  - Forward area $E_x$: the region, as determined by the selected geomasking method, to which a point $x$ is displaced (e.g., a circular ring within donut masking). The set $A'$ restricted to the forward area $E_x$ is denoted as:
    $E_{A'}(x) := \{y \in A' \mid y \in E_x\}$

  - Backward area $E'_{x'}$: the region resulting from the inverse application of the selected geomasking method, starting from an anonymized point $x'$, within which the corresponding original point $x$ is known to be contained. We define:
    $E'_A(x') := \{y \in A \mid y \in E'_{x'}\}$, with participation knowledge
    $E'_B(x') := \{y \in B \mid y \in E'_{x'}\}$, without participation knowledge

  - Obviously holds $x \in E'_{x'}$ and $x' \in E_x$

In the following data attack scenarios, we do not distinguish whether the intruder possesses additional quasi-identifiers beyond the geocoordinates. It is assumed that such additional attributes have already been exploited by the data intruder; e.g. by filtering the data points according to the values of these additional attributes. If an intruder, for instance, knows the age $y$ of a sought-after individual, and the anonymized data does also include the attribute *age*, it is assumed that in a first step the data intruder would filter the dataset by age $y$ and afterwards proceed with the attack scenarios as described below.

In the following typology of data attack scenarios solely the spatial relationships are considered, since the aim of this article is to systematically assess and measure the anonymity provided by geomasking.

**Perspective 1: original address $a_i$ as target attribute**

In this attack perspective, an intruder attempts to use an anonymized dataset to infer the origins of the anonymized addresses in $A'$ in order to identify the corresponding original addresses in $A$ and thereby associate the sensitive attributes of the anonymized records with the individuals living at those addresses (identity disclosure). This attack scenario can be carried out both as a *single attack* (match for a single individual) and as a *database cross match* (Vorgrimler & Lenz, 2003).

An illustrative example: an intruder is interested in individuals listed in the dataset as having a rare disease. If the intruder succeeds in linking the geomasked points back to their original residential addresses, they can disclose the addresses of these individuals. Under worst-case assumptions, this could mean that only one person resides at a given address – making unique re-identification possible. Even if several people live at the same address, the residential location is already associated with a confidential statistical value, which could be further narrowed down through additional knowledge.

Equally problematic is the case of a single-family home in which all members suffer from the same hereditary disease. In such situations, despite the presence of multiple household members, a clear association between address and sensitive attributes can still be established.

These examples demonstrate that – even if the number of individuals at a decrypted address remains unknown – the correct assignment of an anonymized address to its corresponding original address already constitutes a form of de-anonymization and must therefore be strictly avoided.

*<u>Scenario 1.1</u>: original address $a_i$ as target attribute, no knowledge on participation, no knowledge on methods applied*

If an intruder has neither knowledge of participation nor knowledge of methods, he or she may attempt to generate knowledge based on publicly available city maps – for example, through Google Maps. A realistic single attack strategy in this case would be to assign each anonymized address $a_i'$ to the nearest real address from $B$ (all real addresses in the study area).

In a database cross match, the records would be linked in such a way that the total displacement distance between all addresses in $A'$ and $B$ is minimized. This approach represents a classical assignment problem and can be solved using record linkage methods in practice (Lenz, 2006).

*Scenario 1.2: original address $a_i$ as target attribute, knowledge on participation, no knowledge on methods applied*

If an intruder possesses knowledge of participation, a realistic attack scenario can be constructed analogously to Attack Scenario 1.1. The difference is that an anonymized point $a_i{}'$ is not assigned to the nearest real address from $B$ (of which there may be many in the surrounding area), but specifically to the nearest participant address from $A$. This significantly increases the likelihood of success compared to Attack Scenario 1.1.

*Scenario 1.3: original address $a_i$ as target attribute, no knowledge on participation, knowledge on methods applied*

An intruder with expertise in the applied methods could attempt to reverse the anonymization procedure in order to reconstruct the original, non-anonymized addresses. In this case, the methods would be applied in reverse to the anonymized points in $A'$, either to directly recover the original addresses in $A$ or at least to narrow down the areas in which the corresponding original addresses must be located. Without knowledge of participation, the intruder has at least all real residential addresses from $B$ available for potential assignment.

The more knowledge a data intruder possesses about the method – for instance, information about parameter settings – the more precisely they can attempt to reverse the method and restrict the backward area.

*Scenario 1.4: original address $a_i$ as target attribute, knowledge on participation, knowledge on methods applied*

If a data intruder possesses both knowledge of methods and knowledge of participation, they can partially reverse the method as described in data attack scenario 1.3. In addition, the availability of knowledge of participation further increases the risk of correct re-identification, as outlined in data attack scenario 1.2, by reducing the set of potential candidates for assignment.

**Perspective 2: real address $b_i$ or original address $a_i$ as quasi-identifier**

This attack perspective describes the scenario in which an intruder, starting from an address known from $B$ or $A$, attempts to find the corresponding anonymized address. In other words, the intruder seeks to link one or more known addresses - or the individuals residing there - to anonymized addresses from $A'$ and thereby uncover the associated sensitive attributes (attribute disclosure).

This perspective applies, for example, to a data intruder who knows precisely whom they are targeting and where the individuals of interest reside. The intruder then aims to reveal confidential information about these individuals in the anonymized dataset, such as medical records, specific ICD codes, or similar attributes.

Depending on the additional knowledge available to the data intruder, the following attack scenarios can be carried out.

*Scenario 2.1: real address $b_i$ as quasi-identifier, no knowledge on Participation, no knowledge on methods applied*

If a data intruder has neither knowledge on the methods nor knowledge on whether one or more targeted individuals are contained in the anonymized dataset, a possible attack strategy would be to assign the addresses they seek to disclose to the nearest anonymized addresses from $A'$. A database cross match could, as described in scenario 1.1, involve minimizing the total displacement distance between $B$ and $A'$.

In this scenario, however, the intruder does not know whether the targeted individuals or addresses are even included in the dataset at all.

*Scenario 2.2: original address $a_i$ as quasi-identifier, knowledge on participation, no knowledge on methods applied*

This attack scenario is only relevant if the anonymized individuals differ in terms of the expression of a sensitive attribute. If this is not the case, knowledge of participation alone would already allow a data intruder to draw conclusions about sensitive attributes without having to establish a direct link to an anonymized address.

A realistic attack strategy would then consist of assigning the known original addresses from $A$ to the nearest anonymized addresses in $A'$.

*Scenario 2.3: real address $b_i$ as quasi-identifier, no knowledge on participation, knowledge on methods applied*

If a data intruder has knowledge on the methods, he is able to reproduce the masking procedure starting from a known address in $B$ aiming to attack, in order to establish a direct link to an anonymized address in $A'$. The more information the intruder possesses about the functioning of the methods and its parameters, the more precisely he can narrow down the area in which the corresponding anonymized address in $A'$ must lie.

*Scenario 2.4: original address $a_i$ as quasi-identifier, knowledge on participation, knowledge on methods applied*

If a data intruder possesses both knowledge on participation and on the geomasking method being applied, he or she is able to reproduce the masking as described in scenario 2.3. This is done by starting from a known address (in this case in $A$) and establishing a direct link to the perturbed address in $A'$ or at least narrowing down the area in which this anonymized address is located. The participation knowledge provides an intruder with certainty that the address to be linked was part of the target survey.

# 3    Measuring the anonymity

The goal of anonymizing spatial data should always be to safeguard against the data attack scenarios outlined above.

A metric that has been used for decades for evaluating the anonymity of a dataset is the so-called *k-anonymity*. By definition, a dataset is *k*-anonymous if, for any record *r*, there exist at least *k*-1 other records that cannot be distinguished from *r* based on their quasi-identifiers. Consequently, the degree of anonymity increases with larger values of *k*. Accordingly, the probability that a data intruder can correctly identify an original location is *1/k*.

The concept of *k*-anonymity has also been extended to the measurement of anonymity in spatial data and is regarded as the most commonly applied metric in this domain (Broen et al., 2021): *spatial k-anonymity*. This measure describes the number of masked points that are closer to or equally close to the original location than the masked original point itself (Houfaf-Khoufaf & Touya, 2021).

However, the literature demonstrates that *spatial k-anonymity* alone is insufficient for evaluating the anonymity of a geomasked dataset (Allshouse et al., 2010; Steffen et al., 2025; Hauf, 2007). Moreover, different interpretations of how to measure *spatial k-anonymity* have emerged (Broen et al., 2021). For this reason, new metrics will be introduced in the following section that allow for assessing the degree of anonymity of a masked dataset in connection with the attack scenarios outlined in Chapter 2.

### Metric 1: *k-original*

Measures anonymity with respect to data attack scenarios 1.1 and 1.2.

In these two attack scenarios, a data intruder attempts to identify the corresponding original addresses *a* starting from one or several anonymized addresses *a'*. In both cases, the intruder has no knowledge on the methods applied. The difference is that due to scenario 1.2 an intruder possesses participation knowledge.

With the lack of knowledge on the methods, it is most natural to assign the anonymized addresses *a'* to the nearest non-anonymized addresses *b* (without participation knowledge) or to the nearest original addresses *a* (with participation knowledge).

To ensure anonymity under this perspective, it is necessary that for each anonymized address *a'* at least *k*-1 real addresses *b* (or original addresses *a*) are located closer to *a'* than its associated true address.

This can be measured as follows: for each anonymized point *a'*, draw a circle with radius equal to the Euclidean distance to its corresponding original address a, and then count how many real addresses from *B* (without participation knowledge) or original addresses from *A* (with participation knowledge) are lying within this circle.

Without participation knowledge we set:

$$r_i := d(a_i', a_i)$$
$$\mathsf{C}_B\,(a_i', r_i) := \{y \in B \mid d(a_i', y) \le r_i\}$$
$$k_{orig}^{(B)}(a_i') := |\mathsf{C}_B\,(a_i', r_i)|$$

With participation knowledge we set:

$$r_i := d(a_i', a_i)$$
$$\mathsf{C}_A\,(a_i', r_i) = \{y \in A \mid d(a_i', y) \le r_i\}$$
$$k_{orig}^{(A)}(a_i') = |\mathsf{C}_A\,(a_i', r_i)|$$

In Figure 1, *k-original* is determined exemplarily for the anonymized point *a′*. The resulting *k-original* is equal to 3, since the points a, f, and d are located within the yellow circle around a′.

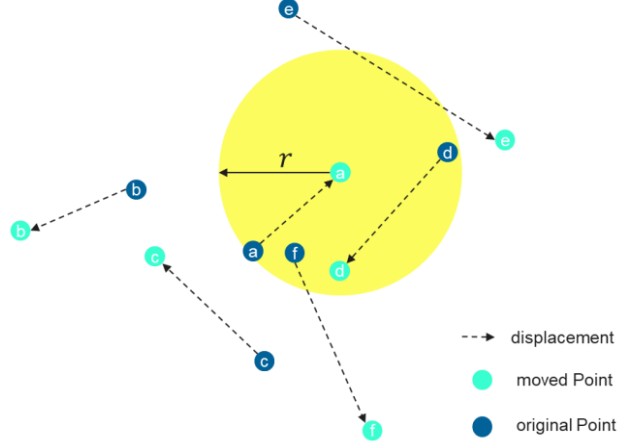

*Figure 1: k-original*

## Metric 2: *k-original method related*

Measures anonymity with respect to data attack scenarios 1.3 and 1.4.

In these two data attack scenarios, a data intruder attempts to identify the associated original addresses *a* starting from one or more anonymized addresses *a′*. In both cases, an intruder is assumed to possess knowledge on the methods applied. The difference is that due to scenario 1.4 an intruder also possesses participation knowledge.

It is natural that the intruder attempts to reverse the known masking method starting from an anonymized address *a′*, insofar as this is possible. To guarantee anonymity in this scenario, the method must not be invertible. Instead, a geomasking method should always contain random components, so that reversing the method results only in some bounded area within which the corresponding original address is located. To ensure anonymity, there must also exist additional candidate addresses within this area.

The metric *k-original method related* therefore determines for each anonymized address *a′*, how many addresses *a* (with participation knowledge) or *b* (without participation knowledge) fall within the area defined by reversing the method.

Without participation knowledge we set:

$$k_{orig_{meth}}^{(B)}(a_i') := |E_B'(x')|$$

With participation knowledge we set:

$$k_{orig_{meth}}^{(A)}(a_i') := |E_A'(x')|$$

**Metric 3: *k-moved***

Measures anonymity with respect to data attack scenarios 2.1 and 2.2.

In these two data attack scenarios, a data intruder attempts to identify the associated anonymized addresses $a'$ starting from one or more original addresses $a$ (with participation knowledge) or real addresses $b$ (without participation knowledge). In both scenarios, the intruder is assumed to have no knowledge on the methods applied. In scenario 2.2 an intruder additionally possesses participation knowledge.

Due to the lack of knowledge on the method being applied, it is natural that the attack would consist of assigning a known address $a$ or $b$ to its nearest anonymized address $a'$. To ensure anonymity under this perspective, it is necessary that for each original address $a_i$ that is indeed part of the anonymized dataset, there exist other anonymized addresses $a'$ that are closer to $a_i$ than its associated anonymized address $a_i'$. If a real address from $B$ is attacked that does not belong to $A$, a correct assignment is by nature excluded. Consequently, the anonymity criterion needs to be verified only for the original addresses $a \in A$, regardless of whether participation knowledge is available.

The criterion can be measured as follows: for each original address $a$, draw a circle with radius equal to the Euclidean distance to its associated anonymized address $a'$, and then count how many anonymized addresses $a' \in A'$ are lying within this circle environment. Here, the associated anonymized address is counted as well.

$$r_i = d(a_i, a_i')$$
$$C_{A'}(a_i, r_i) = \{y \in A' \mid d(a_i, y) \leq r_i\}$$
$$k_{moved}(a_i) = |C_{A'}(a_i, r_i)|$$

In Figure 2, *k-moved* is determined exemplarily for the original point a. The resulting *k-moved* equals 3, since the points a', d', and c' are lying within the yellow circle environment around a.

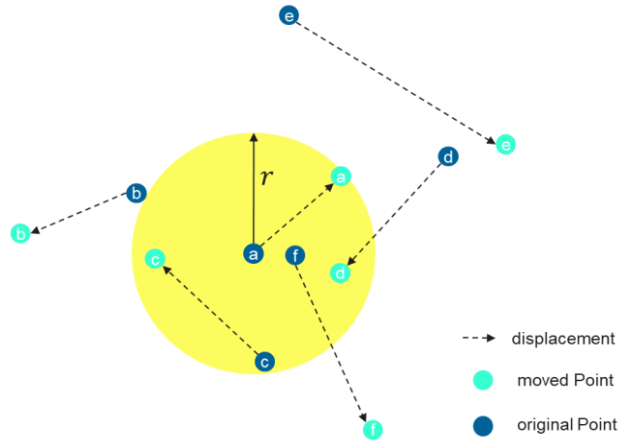

*Figure 2: k-moved*

**Metric 4: *k-moved method related***

Measures anonymity with respect to data attack scenarios 2.3 and 2.4.

In these two data attack scenarios, a data intruder attempts to identify the corresponding anonymized addresses *a'* starting from one or more original addresses *a* (with participation knowledge) or real addresses *b* (without participation knowledge). In both scenarios, an intruder is assumed to possess knowledge on the methods being applied. The difference is that in scenario 2.4 a potential data intruder additionally possesses participation knowledge.

A realistic strategy would be to reproduce the masking method starting from the targeted address *a* or *b*. To guarantee anonymity in this scenario, the method must not be invertible (compare with the explanations in subsection 3.2).

Accordingly, the metric *k-moved method related* measures for each original address *a*, how many anonymized addresses $a' \in A'$ fall within the area defined by reproducing the method. That is,

$$k_{orig_{meth}}^{(A')}(a_i) := |E_{A'}(a_i)|$$

As with metric 2, there is no need to distinguish whether or not an intruder possesses participation knowledge. A successful attack starting from a real address in *B* that is not contained in the original dataset *A* is impossible, since such an address cannot be included in the anonymized dataset *A'*. Hence, measuring anonymity for this case is irrelevant.

## 4 Conclusion

The data attack scenarios and the corresponding metrics for measuring appropriately the spatial anonymity outlined in this paper demonstrate that the approaches used so far for evaluating spatial anonymity of geomasked data are insufficient. The *spatial k-anonymity* commonly employed in literature corresponds to metric 3, introduced above as *k-moved*, and thus measures anonymity solely with respect to the data attack scenarios 2.1 and 2.2. However, since all scenarios sketched in this paper represent realistic attack situations, relying solely on the concept of *spatial k-anonymity* is inadequate for determining the anonymity of a georeferenced dataset. Data holders relying exclusively on *spatial k-anonymity* run the risk that sensitive information may not be sufficiently protected. The authors therefore recommend systematically applying the proposed metrics to assess the required anonymity of geomasked datasets.

Furthermore, in the authors' opinion the predefined minimum threshold of the *k*-value used - particularly concerning the metrics *k-original* and *k-moved* - should not be communicated to the data users. Publishing some chosen *k*-value may itself enable inferences that might weaken anonymity. For example, if it were known that a certain *k* was applied for *k-moved*, a potential data intruder could deduce that the *k*-1 nearest anonymized neighbors do not represent candidates for a correct assignment. This would, in turn, increase vulnerability to data attacks.

Future work should examine the practical application of the metrics proposed here.